\documentclass[printer]{tPHM2e}
\usepackage{graphicx}

\begin{document}
\doi{10.1080/1478643YYxxxxxxxx}
 %\issn{1478-6443}
%\issnp{1478-6435}  \jvol{86} \jnum{9} \jyear{2006} \jmonth{21 March}

\markboth{Modelling quasicrystals}{A. Losev}

\title{Wave models of non-crystallographic structures}

\author{A. LOSEV \thanks{$^\ast$ Email: alosev@svr.igic.bas.bg}$^\ast$\\\vspace{6pt} 
Institute of General and Inorganic Chemistry,
G. Bonchev str., bldg 11, Sofia 1113, Bulgaria\\\vspace{6pt}
\received{v2 November 2007}}

\maketitle

\begin{abstract}
Combinations of oscillations with incommensurate periods are utilized to generate discrete planar structures with non-crystallographic symmetries. Some previously known tilings and results are reconsidered in this context. A new kind of decagonal tiling, its relations with the Penrose tilings and some experimental reports are discussed.
\end{abstract}

\section{Introduction}
 Twenty five years after their discovery\cite{Shecht},  quasicrystals have become an accepted object of academic research. The  existence of nonperiodic structures with long range order came as a surprise for most of the physics community, even if in mathematics the subject had already been explored. Different formal descriptions have been proposed and shown to be more or less equivalent. Substitutions, matching rules, projection and cuts from higher dimensions, grid methods, coverings, nonlinear dynamics are approaches which have dealt successfully with various aspects of the topic. Physical phenomena which exhibit nonperiodic order generate further insight into the subject. A simple experiment demonstrated how a wave pattern with dodecagonal symmetry \cite{Edw} arises in fluids subjected to forced oscillations with just two frequencies.  Other quasiperiodic systems of Faraday waves appear to be possible\cite{Binks} but earlier  setups failed to produce either the eightfold or tenfold cases. The idea has evolved  evolved naturally in optics where various findings\cite{Steurer} have been reported. A recently study\cite{Gorkhali} generated results for oddnumbered  cases.
Looking for the origin of aperiodic structures in incommensurate periods can be done in various ways. Abstract pair potentials are able to reproduce the symmetry of an initial atomic configuration and allowed to reproduce\cite{LM1,LM2} a perfect 2D octogonal Ammann-Beenker tiling by considering the  superposition of periodic waves. Studies with the Dzugutov potential have been already shown to obtain dodecagonal quasicrystals\cite{Dzugu}. An other viewpoint is offered by nonlinear dynamics where it was observed that chaos may develop as structures with noncrystallographic symmetry\cite{Zas,Low}.
The first observed quasicrystal was reported as the three-dimensional structure of an alloy but soon it became apparent that aperiodicity can be confined in fewer dimensions as quasicrystals with planar decagonal, octogonal and dodecagonal\cite{Bender, Wang, Ishimassa} symmetry were observed. Tilings of these types were obtained as coverings which were shown to possess also extremal properties\cite{Steinh, Shlomo, Gahler-max, Gumm-}. The distinction between geometry and physics became blurred and even  higher-dimensional models began to be viewed realistically. 

Relying on incommensurability, below we outline a variant approach and reconsider briefly some known tilings.  A new instance of a decagonal tiling, its relations with  the Penrose tilings (PT) and some experimental reports is discussed.

\section{Model and Method}
The vertices of a tiling form the discrete set of points ${\bf r_k}$ which admits a representation of the type
${\bf r_k}=\sum_i^na_{ki}{\bf e_i}$
where ${ {\bf e_i} }$ is a basis set of $n$ vectors and $a_{ki}$ are appropriately chosen integers. It is a remarkable fact that a quasiperiodic lattice can be obtained with  a set of unit vectors pointing to the vertices of a polygon centered at the origin 
and integers $a_{ki}$ selected by some simple rules. This is indeed the gist of the cut and project method but this rather general formula allows other readings. Studies of nonlinear phenomena have shown \cite{Zas, Low} that the quasiperiodic resonance Hamiltonian for an oscillator subjected to perturbations can be recast in the general form 

$ H(u,v)=A\sum_{k=1}^n \cos{\xi_k}$

where $\xi_k=v \cos{({2\pi \over q} k)}+u \sin {({2\pi \over q} k)}$.  This can be rewritten in compact form as $\xi_k= (u,v). {\bf e_k }$. Mapping the density of points in $u, v$ space, a 'web' with a quasicrystalline fractal structure is observed. The result is in accordance with the  general Landau symmetry principle. Levy and Mercier\cite{LM1,LM2} arrived at essentially the same general result by considering the density distribution induced by pair potentials. These results suggest an  interpretation which would consider the extrema of a continuous function, their distribution in space and neighbor linking. As both types of these studies noted, an initial configuration propagates with almost exact copies occurring at nearly equivalent locations. 

Taking a cue from these results we chose to investigate a sum of coupled oscillations of the type 

 $z(x,y)=\sum _k A_k cos(\alpha_k x) cos(\beta_k y)$. 

where the coefficients $\alpha, \beta$ and the number of terms reflect
symmetries. Assuming that a discrete structure is hidden within the continuous one, a discretizing strategy should be implemented. It goes without saying that this is the crux of the method. The finite step chosen for the horizontal $x, y$ variables should allow a sufficiently fine graining of $z$. Looking for the configuration outlined by some particular constant value $z_0$ hardly reveals noteworthy structures, but the result improves significantly if some weaker criterion is used. For instance, performing a discretization of $z$ by considering only non-zero values of $round(z)$ or $floor(z)$ already allows to reproduce some interesting results. Imposing a threshold for $z$, i.e. $z>z_0$, allows a finer tuning of the selection. A small step in the horizontal plane ensures that no occurrences of $z>z_0$ are missed but tends to generate clustered points which blur the features of the structure. To make it more evident links between points $r_1, r_2$ lying at a distance within an interval are drawn i.e. two points are linked if $L-d<\vert r_1-r_2 \vert <L+d$. The value of $L$ is chosen from geometrical considerations. The tolerance measured as $d/L$ is not a sensitive parameter, as long as it not zero.  Starting with a threshold $z_0$ close to the absolute maximum of $z$ obtains just a few points. Decreasing its value adds more points to the selection. The structures are seen to emerge, barely outlined at first, with more features gradually added. When a larger element allows multiple decompositions, the variants tend to manifest themselves all at the same time, revealing locations for phason flips.
Indeed this approach can be seen as a literal interpretation of the grid method from which it is known that vertices occur in the vicinity of multiple intersections\cite{Steurer}.

\section{Results and Discussion}
\subsection{Twelve- and Eight- fold cases}
A dodecagonal aperiodic tiling was constructed first by the grid method from two regular hexagonal meshes rotated at $\pi/6$ \cite{Stampfli}. Later, variants known as 'shield tiling' were obtained and the general case was elucidated by the cut-and-project method \cite{Gahler12}. However  the existence of twelvefold  wave patterns in fluids lead to  reconsiderating\cite{Arbell} the structure as arising in direct physical space. Using

$z=1/2 \cos(a x)+1/2 \cos(a y))+2\cos(b x)\cos(a y/2)+
2\cos(a x/2)\cos(b y)$,

where $a=2\pi$, $b=2\pi\cos(\pi /3)$ and setting a threshold $z_0=4.6$ allows to obtain easily patches of such a tiling as shown on Fig.1. Larger patches contain 'defects' which suggest to look for some more sophisticated procedure if obvious corrections by direct intervention are to be excluded. 
\begin{figure}
\resizebox{100 mm}{60 mm}{ \includegraphics{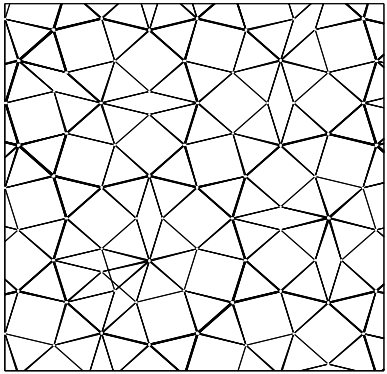}}
\caption{The twelvefold structure emerging above $z_0=4.6$: a patch of dodecagonal tiling containing one superfluous point.}
\label{Fig.1}      
\end{figure}
However geometrical perfection may be considered an artefact in a  physical system and some degree of imperfection would be indeed a realistic feature. This introductory example only serves as a demonstration of the method proposed here. The expression for $z$ allows to consider variants for grouping the terms of the sum which alternatively are seen as different superpositions of waves. Modifying the weights, or 'amplitudes', and the threshold offers some insight about the nature of the model. 
The eightfold case can be also reproduced with this method but it has been already been exposed  at some length in a study developing a  somewhat similar approach\cite{LM1,LM2}.

\subsection{The Ten-fold case}
Proceeding just as in the previous cases fails to obtain directly the expected result. However investigating the particular instance of an incomplete sum produces an interesting variation as seen on Fig.2. It has been obtained by retaining just two terms with equal amplitudes and  $\alpha_1=2\pi \tau$ and $\alpha_2=\alpha_1 \tau^{-2}$, where  $\tau=2\cos({\pi \over 5})$ which is the golden mean, and $\beta_1=\sin{({4\pi \over 5})}, \beta_2 = \sin{({2\pi \over 5})}$. The threshold has been set to $z_0=1.65$, the basic length is $L=1.8$ with a tolerance $d=0.14$, the result being plotted with the origin in the lower left quadrant. 
\begin{figure} [! h b p]
\resizebox{100 mm}{70 mm}{ \includegraphics{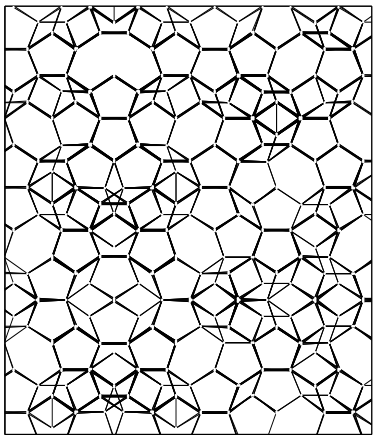}}
\caption{The tenfold structure emerging above $z_0=1.65$, plotted with the origin in the lower left quadrant.}
\label{Fig.2}      
\end{figure}
A decagonal motif consisting of two pentagons, three thin rhombuses and an elongated hexagon is clearly present. As secondary features one might note crossed stars, dismantled hexagons and other complications. Setting a higher threshold filters most of them out. Retaining only the most regular elements of this structure, it is possible to devise a geometrical reconstruction as the one shown in Fig.2. It has been obtained by drawing tessellated decagons which overlap only over thin rhombuses, following the outlines given by the discretization of $z$. Some ambiguities have been resolved by the decision to preserve pentagons, allowing the decagons eventually to cut across some non-pentagonal elements. It was realized that a better template would be a pentagonal Penrose tiling (P1). Both tilings are built with pentagons, rhombuses and some additional shapes which  allow  interrelated transforms. When a pentagon is inscribed within a hexagon, the crown shaped fragment, which remains, can be reunited with a rhombus to form a 'boat' and, further, a boat and a crown can form a star. Such transformations work both ways and in fact Fig.3 is locally equivalent to a patch of P1 around a center of tenfold symmetry.   Figure 4 corresponds  to a covering around the global fivefold center of P1. Either as an autonomous structure or as a covering of P1, this construction presents an  alternative, intriguing by its simplicity and flexibility.

It appears that such  tilings can be obtained with a single figure when some rules are observed. The basic element  is a decagonal ring with ten protruding 'spikes', which suggest to adopt the descriptive name 'spiky decagon tiling' (SDT). Of course the spiky decagon is not exactly a tile, just as 'covering' is not the same as 'tiling'. However the spikes produce a tessellation both inside and outside decagons and a tiling with three shapes is obtained.  Isolated, i.e. non-overlapped, decagons also appear but then a similar tessellation is possible in ten different orientations. Five incompletely tessellated decagons appeared close to the global center in Fig.4, with two alternatives for each; preserving the symmetry leaves just two general options. Obviously, as a geometrical construct the SDT remains  underdetermined which would reflect the omitting of one term in the sum for $z$. When the P1 is transformed into  a SDT some information is lost and, conversely, additional data is needed to build a SDT equivalent to a Penrose tiling. A pentagon can be cut from either side of the hexagons in SDT and if this is performed in a disordered  way, when crowns are reunited with other elements, a  variant of the random pentagonal Penrose tiling \cite{Gahler-max,Gumm-,Reich} would be produced. Here a  constitutive rule of the SDT is made apparent, namely that hexagons should not appear on opposite sides of a shared rhombus. Otherwise twisted stars would appear in reconstruction and more generally, trivial periodicity would be possible.
\begin{figure}
\resizebox{200 mm}{125 mm}{  \includegraphics{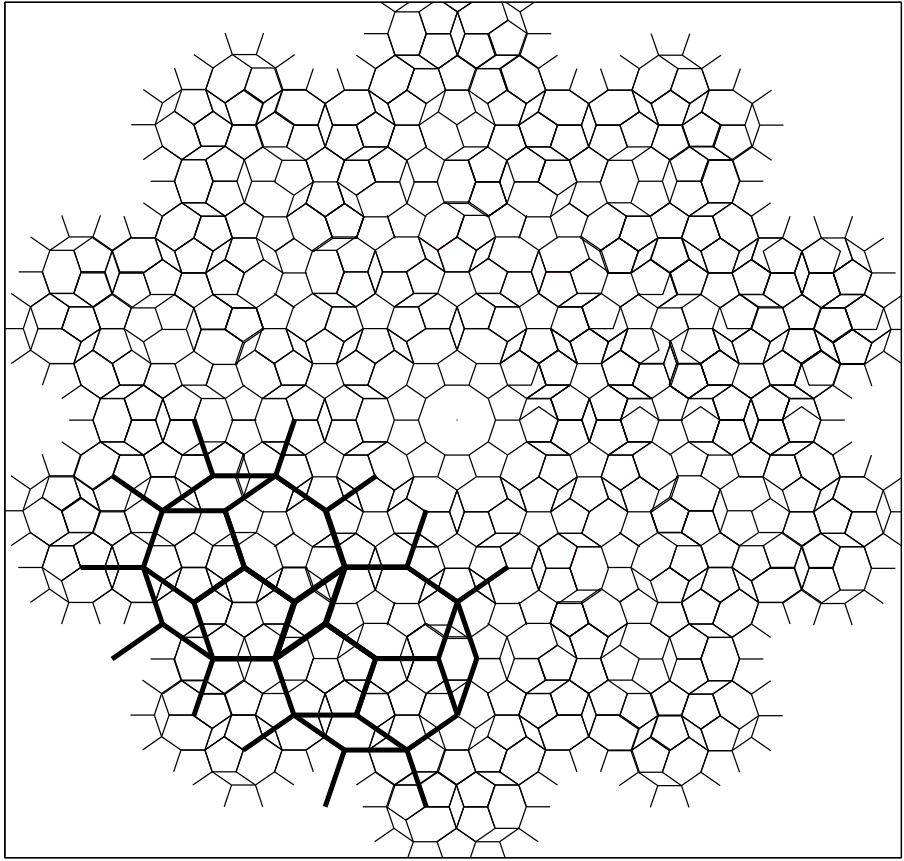}}
\caption{The Spiky Decagon Tiling. Equivalence with P1 around a center of tenfold symmetry is shown within the first $\pi/10$-sector and the hierarchical structure is outlined {\bf (in bold)} in the lower left sector.}
\label{Fig.3}      
\end{figure}
 As derived from a superposition of incommensurate waves a SDT appears to be a nonperiodic decagonal tiling which stands on its own. Extending it by local steps, which preserve the tenfold symmetry, proceeds by forced or unforced moves, just as in a PT, and no obstacle for an indefinite growth has been found.  A simple inflation scheme shows that the decagonal structure of the tiling could be extended indefinitely, if successive pairs of rings made from pentagons and hexagons are used, the enlargement at each step being $2\tau+1$. Inflating a tessellated decagon also appears to be an option, as shown on Fig.3, but the details for the three tiles have to be worked out. A similarity with the P1 is also retained in this respect. Under inflation the pentagons in P1 are known to be of three different kinds. In the covering SDT one kind never appears inside decagons while the other two, which are actually present, are from different kinds; the one obtained by cutting the hexagon is of either kind, depending on which side the crown is taken out. The rules to obtain it as an aperiodic tiling, still  need to be elucidated. Equivalence with P1 is an open possibility as the partial reconstruction on Fig.3 and the derivation of Fig.4 suggest. Successful reconstructions of the PT have exploited two main lines of approach: decagons or golden triangles. Gummelt's decagons\cite{Gumm} produce a perfect PT if  just two kinds of overlap are allowed. With more relaxed rules\cite{Gahler-max,Gumm-,Reich} a construction rather similar to the one  described here is obtained. On the other hand, investigating a covering of a golden triangles tiling, Kasner and Papadopols\cite{Kasner} have arrived at the same uniformly overlapping decagons comprising the elongated hexagon, two pentagons and three rhombuses. There is however some difference with their result as arcs

\begin{figure}
\resizebox{125 mm}{75 mm}{ \includegraphics{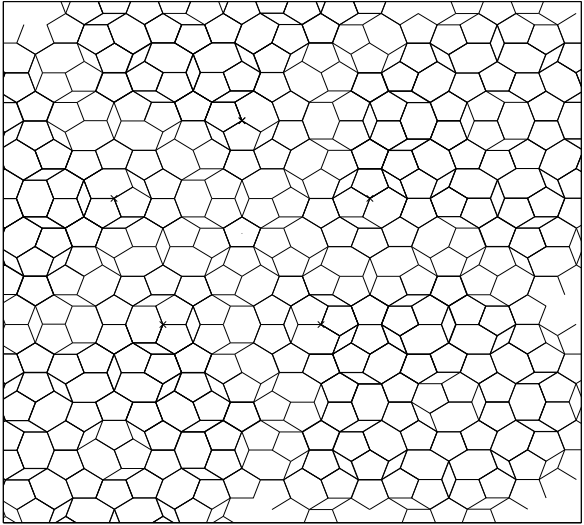}}
\caption{The spiky decagon covering around the global fivefold center of P1.}
\label{Fig.4}      
\end{figure}

with two or seven  hexagons are present in their tiling and not with only four as in Fig.3 and 4. A subtle difference between the underlying structures\cite{Baake} is known to exist. Now, with hindsight, a modest patch of SDT can be recognized also in a work which investigated magnetisation energy on a T\"ubingen triangle tiling\cite{Grimm} and it is expected to be of the Kasner-Papadopolos variety. An unambiguous patch of SDT such as the one seen on Fig.2 has been obtained by optical means as a photonic crystal\cite{Gorkhali}. The setup to produce it relies on laser beams emanating from the vertices of a pentagon and crossing in a single point. The authors describe the resulting structure (shown in their Fig.8) as a Penrose tiling. Most physical structures appear to be rather irregular and in fact look more like Fig.2 rather than with the geometrical reconstruction of Fig.3. If it is possible to cut crowns in a disorderly way there might be also a possibility to cut two of them from the same hexagon, leaving just a fat Penrose rhombus. It is perhaps worth noting that a projection of the icosahedron can produce an elongated hexagon, where the outer vertices appear just at the points which decompose it into a fat rhombus and two crowns\cite{Gratias}. But indeed the second crown is cut from a pentagon. This hints at a transition from a pentagonal Penrose tiling (P1) to a rhombic one (P3) on the same scale. Often experimental images from quasicrystals with decagonal or icosahedral symmetry also exhibit the same untidy arrangements \cite{Reich, Led} which are strongly reminiscent of the initial look of the SDT as seen on Fig.2, the complications and geometrical ambiguities being present ab initio and not as additions. 

A particularity of the SDT is that only convex tiles are actually present. Physically the tiles have less importance than the vertices. The basic configuration of the spiky decagon has only one type of threefold vertex. In the tiling, points where the decagons cross each other, produce fourfold vertices. So the tiling has only two types of vertices. The overlap introduces also a new length scale with the appearance of the rhombus and its short diagonal, which marks two vertices closer than the unit length and yet unconnected. However the tessellation of a decagon is still done only with the unit length. This is another simplification of Gummelt's decagons whose tessellation is done with overlapping concave kites which already have sides of different lengths. The simplified overlapping is traded for complication in the exact rules which govern the construction of nontrivial tilings.

\section{Conclusion}
Continuous models for quasicrystals offer a most natural way to account for imperfection in structures that are conceived as ideal models.  Constructive interference is the obvious example of discrete patterns arising from oscillations. In a simplistic view waves with incommensurate periods would be unlikely to produce interesting patterns. However the concept of quasiperiodic function in two or more dimensions offers a view on a more complicated reality, just as quasicrystals revealed new riches in the world of crystals. A development of such an  approach has allowed to reproduce some known results and suggested a new kind of tiling which appears to be a member of the D\"urer-Kepler-Penrose \cite{Luck} family.

\label{lastpage}

\end{document}